\documentclass[aps,prl,twocolumn,showpacs,floatfix,amsmath]{revtex4}

\usepackage{graphicx}
\usepackage{hyperref}
\usepackage{bm}

\newcommand{\qsys}{{\cal Q}}
\newcommand{\bfB}{\mathbf{B}}

\newcommand{\be}{\begin{equation}}
\newcommand{\ee}{\end{equation}}
\newcommand{\bea}{\begin{eqnarray}}
\newcommand{\eea}{\end{eqnarray}}

\begin{document}
\title{Noisy Control, the Adiabatic Geometric Phase, and Destruction of 
the Efficiency of Geometric Quantum Computation}

\date{\today}

\author{Frank Gaitan}

\email{gaitan@physics.siu.edu}

\affiliation{Department of Physics; Southern Illinois University;
              Carbondale, IL 62901-4401}

\begin{abstract}
We examine the adiabatic dynamics of a quantum system coupled to a noisy
classical control field. A stochastic phase shift is shown to arise in the
off-diagonal elements of the system's density matrix which can cause 
decoherence. We derive the condition for onset of decoherence, and identify 
the noise properties that drive decoherence. We show how this 
decoherence mechanism causes: (1) a dephasing of the observable consequences 
of the adiabatic geometric phase; and (2) the loss of computational efficiency 
of the Shor algorithm when run on a sufficiently noisy geometric quantum 
computer. 
\end{abstract}


\maketitle

As the adiabatic geometric phase approaches its twentieth birthday 
\cite{bry,gph} it continues to impact research in physics and chemistry.
The original scenario considered a quantum system $\qsys$ with a discrete,
nondegenerate energy spectrum whose dynamics was driven by a classical set of
control parameters $\mathbf{B}(t)$. It was shown that when $\mathbf{B}(t)$ 
was cycled adiabatically
through a loop in parameter space, and $\qsys$ was initially prepared in an
eigenstate $|E(0)\rangle$ of the initial Hamiltonian $H(0)$, its quantum state 
would return to the initial state $|E(0)\rangle$ at the end of the cycle,
to within a phase factor $\exp [i\phi ]$. The phase shift $\phi$ contained
a geometric contribution now known as the adiabatic geometric 
phase (AGP) which had been discarded in prior treatments of quantum adiabatic 
dynamics. Since then, this original scenario has been generalized in a number 
of ways: (1) the requirement of nondegeneracy was removed \cite{wze}; (2) the
adiabatic restriction was removed \cite{AhA}; (3) the control field
$\mathbf{B}(t)$ was allowed to have its own dynamics \cite{dyn}; and 
(4) $\qsys$ was allowed to interact with a quantum environment as well as
with the control field $\mathbf{B}(t)$ \cite{qe1,qe2,qe3}. In this Letter we 
return to the original scenario, though we allow the classical control field
to contain a noise component $\mathbf{B}_{n}(t)$ along with the 
(deterministic) adiabatically varying component $\mathbf{B}_{a}(t)$:
$\mathbf{B}(t) = \mathbf{B}_{a}(t) + \mathbf{B}_{n}(t)$. 
We will show that $\bfB_{n}(t)$ causes a stochastic phase shift to appear in 
the off-diagonal elements of $\qsys$'s density matrix which, for sufficiently
strong noise, causes decoherence. We derive the condition that determines the 
onset of decoherence (see eq.~(\ref{deco2})), and identify 
the noise properties that drive the decoherence. \textit{We show that this 
decoherence mechanism causes all physical consequences of the AGP to become 
unobservable}. 

The AGP has also had an impact on the new research area of quantum computation.
In 2000, a proposal was put forth for a geometric quantum computer (GQC) 
\cite{gqc} which uses the AGP to encode conditional phase shifts into the 
quantum state of a GQC. It was argued that the AGP gave the GQC
a modest degree of fault tolerance to uncertainty in the control parameters, 
though the need to examine the robustness of a GQC to decoherence was also
noted. In the second part of this Letter we carry out such a decoherence 
analysis. We show that the above decoherence mechanism arising from noisy
control severely impacts the performance of a quantum algorithm run on a noisy
GQC. Specifically, we show how noisy control causes the Shor algorithm to
loose its computational efficiency so that the time to factor an integer
($N$) grows exponentially with the size of the integer ($\log N$). We 
establish the condition for the onset of decoherence, and show that the success 
probability for the algorithm is smoothly degraded as the noise becomes 
stronger, reaching the point at onset where the computational efficiency of 
the Shor algorithm has been completely destroyed. We also show that at 
onset of decoherence, the entanglement fidelity of a 2-qubit entangled state
is degraded to that of a mixed state upon output from a sufficiently noisy
controlled phase gate of the type used in a GQC. This 
noise-induced loss of entanglement confirms that the GQC is no longer behaving 
as a quantum computer. To the best of our knowledge, \textit{this Letter 
provides the first quantitative demonstration of how the computational 
efficiency of a landmark quantum algorithm such as Shor's is destroyed when it 
is run on a sufficiently noisy GQC}. A detailed presentation of all arguments 
will be reported elsewhere \cite{fg3}.

\textit{1. General Analysis:} We consider the situation where the dynamics of
an N-qubit system $\qsys$ is driven by an external control field $\bfB (t)$.
The control field is composed of a deterministic component $\bfB_{a}(t)$
and a noise component $\bfB_{n}(t)$. We will be interested in cases where 
$\bfB_{a}(t)$ executes a cyclic evolution with period $T$. We take the noise 
component $\bfB_{n}(t)$ to be a stationary stochastic process with
zero mean ($\overline{\bfB_{n}(t)} = 0$) and noise correlation time $\tau_{c}$.
To simplify the notation we assume the noise is isotropic. This assumption is
not essential, and the following analysis can easily be repeated without it.
For isotropic noise, the noise correlation function 
$\overline{B_{n}^{i}(t)B_{n}^{j}(t^{\prime})} = \delta_{ij}\sigma^{2}f(\tau )$.
Here: (1) $\delta_{ij}$ is the Kronecker delta;
(2) $\sigma^{2}$ is the variance of the noise $\overline{(B_{n}^{i}(t))^{2}}$;
(3) $\tau = t - t^{\prime}$; and (4) $f(\tau )$ is the normalized noise 
fluctuation profile with peak value $f(0) = 1$ and temporal width of order 
$\tau_{c}$. Because of the stochastic nature of 
$\bfB_{n}(t)$, each application $k$ of the control field contains a different 
realization of the noise $\bfB_{n}(t;k)$. 
To account for the dynamical effects of noise, it is necessary to
introduce an ensemble of identical quantum systems $\qsys_{k}$, all prepared
in the same initial state $|\psi (0)\rangle$, but whose evolution is subject 
to a different noise realization $\bfB_{n}(t;k)$.

Quite generally, the system Hamiltonian $H(t)$ can be written as the sum of 
two terms: $H_{a}(t)$ and $H_{s}(t)$. Here $H_{a}(t)$ is the Hamiltonian in 
the absence of noise and depends solely on $\bfB_{a}(t)$. The stochastic 
interaction term $H_{s}(t)$ depends explicitly on the noise component 
$\bfB_{n}(t)$ and its action varies from one application of the control field 
to another. We introduce the instantaneous energy 
eigenvalues $E_{n}(t)$ and eigenstates $|E_{n}(t)\rangle$ of the noiseless 
Hamiltonian $H_{a}(t)$: $H_{a}(t)|E_{n}(t)\rangle = E_{n}(t)|E_{n}(t)\rangle$. 
Note that the periodicity of $\bfB_{a}(t)$ implies that $H_{a}(t+mT) = 
H_{a}(t)$. As in the original AGP scenario, the instantaneous energies 
$E_{n}(t)$ are assumed to be discrete and nondegenerate for all $t$ of 
interest so 
that $|E_{n}(t+mT)\rangle =|E_{n}(t)\rangle$. Introducing the minimum energy
level separation $\Delta = \min |E_{n}(t) - E_{m}(t)|$, the noiseless dynamics 
will be adiabatic if $\hbar /T \ll \Delta$. Furthermore, as shown in 
Ref.~\cite{qe1}, the dynamics associated with the noise will also be
adiabatic if $\hbar /\tau_{c} \ll \Delta$. Both of these conditions are assumed
in the remainder of this Letter. 

As noted above, each ensemble element $\qsys_{k}$ sees a different noise
realization $\bfB_{n}(t;k)$ added to $\bfB_{a}(t)$. The dynamics of
$\qsys_{k}$ is driven by $H(t;k) = H_{a}(t) + H_{s}(t;k)$ which generates the
final state $|\psi (t;k)\rangle = U(t_{f},t_{0};k)|\psi (0)\rangle$, 
where $U(t_{f},t_{0};k) = \exp\left[-(i/\hbar)\int_{t_{0}}^{t_{f}}dt H(t;k)
\right]$. Thus the final state produced by the noisy control field varies over 
the ensemble. Consequently, calculation of the expectation value of an 
observable requires an ensemble average over the noise, along with the usual 
quantum mechanical averaging.

We divide the time interval ($t_{0}$, $t_{f}$) into $J$ subintervals of
duration $\epsilon = (t_{f}-t_{0})/J$ by introducing intermediate times
$t_{j} = t_{0} + j\epsilon$ ($j=0, \ldots , J$). Eventually we let 
$J\rightarrow \infty$. The propagator $U(t_{f},t_{0})$ factors into a product 
of propagators: $U(t_{f},t_{0})=U(J)\cdots U(0)$, where 
$U(j) = U(t_{j}, t_{j-1})$. One can show \cite{fg3,fg4} that in the adiabatic 
limit the matrix 
elements $U_{lm}(j) = \langle E_{l}(t_{j})| U(j) |E_{m}(t_{j-1})\rangle$ are:
\begin{eqnarray}
\lefteqn{\hspace{-0.20in} U_{lm}(j)  = } \nonumber \\
 & &  \hspace{-0.2in}\delta_{lm}\left[ 1 - \frac{i\epsilon}{\hbar}
          \left\{ E_{l}(t_{j}) - \hbar\dot{\gamma}_{l}(t_{j}) +
                  H_{s}^{ll}(t_{j})\right\} 
                    + \mathcal{O}(\epsilon^{2})\right] . 
\end{eqnarray}
Here $\dot{\gamma}_{l}(t_{j}) = i\langle E_{l}(t_{j})|\dot{E}_{l}(t_{j})
\rangle$ is the time derivative of the AGP for energy level $l$; 
$\Gamma_{lm}(t_{j}) = \langle E_{l}(t_{j})|\dot{E}_{m}(t_{j})\rangle$ is the 
non-adiabatic coupling for levels $l$ and $m$;
and $H_{s}^{ll}(t_{j}) = \langle E_{l}(t_{j})|H_{s}(t_{j})|E_{l}(t_{j})
\rangle$. Choosing $|\psi (t_{0})\rangle = |E_{k}(t_{0})\rangle$;
reconstructing $U(t_{f},t_{0})$ from the $U(j)$; and letting $J\rightarrow
\infty$ gives:
\begin{equation}
|\psi (t_{f})\rangle = \exp\left[ -i\left\{ \Gamma_{a}(k) + \Gamma_{s}(k)
                            \right\}
                       \right] |E_{k}(t_{0})\rangle . \label{fstate}
\end{equation}
Here $\Gamma_{a}(k) = (1/\hbar )\int_{t_{0}}^{t_{f}} dt \left\{ E_{k}(t) -
\hbar\dot{\gamma}_{k}(t)\right\}$ and $\Gamma_{s}(k) = (1/\hbar )
\int_{t_{0}}^{t_{f}} dt\, H_{s}^{kk}(t)$. Eq.~(\ref{fstate}) shows that
noise introduces a \textit{stochastic phase shift} $\Gamma_{s}(k)$ into the 
final state $|\psi (t_{f})\rangle$ which is sensitive to the initial energy 
eigenstate $|E_{k}(t_{0})\rangle$ through $H_{s}^{kk}(t)$.

To bring out the physical consequences of the stochastic phase $\Gamma_{s}(k)$ 
we consider an initial superposition state $|\psi (t_{0})\rangle = 
\sum_{k} c_{k}|E_{k}(t_{0})\rangle$. From the linearity of 
quantum mechanics: $|\psi(t_{f})\rangle = \sum_{k} c_{k}U(t_{f},t_{0})
|E_{k}(t_{0})\rangle$. Information about the phase coherence of the final 
superposition is carried in the off-diagonal elements of the final density 
operator $\rho (t_{f}) = |\psi (t_{f})\rangle\langle\psi (t_{f})|$. 
One can show \cite{fg3} that the noise averaged density matrix elements are:
\begin{equation}
\overline{\rho_{kj}(t_{f})} = c_{k}c_{j}^{\ast}\exp\left[ -i\Gamma_{a}(k,j)
\right] \overline{D(k,j)} , \label{offdiag}
\end{equation}
where $\Gamma_{a}(k,j) = \Gamma_{a}(k) - \Gamma_{a}(j)$, 
and the decoherence factor $\overline{D(k,j)}$ is:
\begin{equation}
\overline{D(k,j)} = \overline{\exp \left[ -i\Gamma_{s}(k,j)\right]} ,
   \label{Djk}
\end{equation}
with $\Gamma_{s}(k,j) = \Gamma_{s}(k) - \Gamma_{s}(j)$. 
We see that the decoherence factor $\overline{D(k,j)}$ is a direct consequence 
of the noise-induced phase shift $\Gamma_{s}$. Often, $H_{s}(t)$ is
linear in the noise component $\bfB_{n}(t)$ so that $H_{s}(t) = -(\gamma\hbar
/2)\bfB_{n}(t)\cdot\hat{\mathbf{O}}$. Here $\gamma$ is the coupling constant 
and $\hat{\mathbf{O}}$ is a vector operator. In the adiabatic limit, $\tau_{c}
\ll (t_{f}-t_{0})\equiv \mathcal{T}$. Thus, if we partition the integration 
interval ($t_{0}$, $t_{f}$) that appears in the definition of $\Gamma_{s}(k,j)$ 
into $M = \mathcal{T}/\tau_{c}$ subintervals of duration $\tau_{c}$, we render
$\Gamma_{s}(k,j)$ into a sum of uncorrelated random variables 
$\Gamma_{s}^{m}(k,j)$: $\Gamma_{s}(k,j) = \sum_{m=1}^{M}\Gamma_{s}^{m}(k,j)$.
Because the noise is stationary, the $\Gamma_{s}^{m}(k,j)$ have identical
probability distributions. If the $\Gamma_{s}^{m}(k,j)$ are not only 
uncorrelated, but also statistically independent, it follows from the 
Central Limit Theorem that $\Gamma_{s}(k,j)$ will have a Gaussian probability
distribution. In this case, one can show \cite{fg3} that 
$\overline{\Gamma_{s}(k,j)} = 0$ and
\begin{equation}
\overline{\Gamma_{s}^{2}(k,j)} = \frac{\eta\gamma^{2}\sigma^{2}}{4}I_{kj} ,
\label{variance}
\end{equation}
where: (1) $\eta$ cycles of the control field have been
applied ($\eta T = \mathcal{T}$); (2) $I_{kj}=\int_{0}^{T} dt dt^{\prime}
\hat{\mathbf{O}}_{kj}(t)\cdot\hat{\mathbf{O}}_{kj}(t^{\prime})
f(t-t^{\prime})$, with $\hat{\mathbf{O}}_{kj}(t) = \hat{\mathbf{O}}_{kk}(t) - 
\hat{\mathbf{O}}_{jj}(t)$, and $\hat{\mathbf{O}}_{kk}(t) = \langle E_{k}(t)|
\hat{\mathbf{O}}|E_{k}(t)\rangle$; and (3) $f(t-t^{\prime})$ is the normalized
noise fluctuation profile introduced earlier. Having the mean and variance 
of the Gaussian probability distribution for $\Gamma_{s}(k,j)$, we can 
evaluate the noise average in eq.~(\ref{Djk}). The result is
$\overline{D(k,j)} = \exp [ -\overline{\Gamma_{s}^{2}(k,j)}/2 ]$.
One expects that when the phase uncertainty 
$\sqrt{\overline{\Gamma_{s}^{2}(k,j)}}\sim 2\pi$, the superposition in
$|\psi (t_{f})\rangle$ will have been dephased/decohered by the noise. 
Inserting $\overline{\Gamma_{s}^{2}(k,j)}\sim (2\pi )^{2}$ into our 
expression for $\overline{D(k,j)}$ gives 
$\overline{D(k,j)}\sim 3\times 10^{-9}$ so that the off-diagonal elements of 
$\overline{\rho (t_{f})}$ (see eq.~(\ref{offdiag})) are effectively zero and 
the noise has in fact
caused an effective collapse of the wavefunction $|\psi (t_{f})\rangle$. One 
can show \cite{fg3} that the noise variance $\sigma^{2}$, the average noise 
power absorbed by $\qsys$ per unit volume $\overline{P}/V$, and the effective 
bandwidth $\Delta\omega$ of the absorbed noise power are related: 
$\overline{P}/V = \sigma^{2}\Delta\omega$. Using this relation, together with 
eq.~(\ref{variance}), one can re-express the condition for onset of 
decoherence $\sqrt{\overline{\Gamma_{s}^{2}(k,j)}}\sim 2\pi$ as
\begin{equation}
\frac{\eta}{16\pi^{2}}\left(\,\frac{\gamma^{2}}{\Delta\omega}\,\right)
  \left(\, \frac{\overline{P}}{V}\,\right)\, I_{jk} \sim 1 .
\label{deco2}
\end{equation}
We consider a number of applications of this general analysis in the remainder 
of this Letter.

\textit{2. Dephasing the AGP:} In our first application we reconsider the
original AGP scenario \cite{bry}, allowing for noisy control. Here \qsys\ 
is a single qubit, and $\bfB_{a}(t)$ precesses about the z-axis with period $T$
at an angle $\theta_{0}$. Let $C$ denote the contour traced out by the tip of
$\hat{\bfB}_{a}(t)$ in a time $T$. To observe the AGP, the initial state must 
be a superposition: $|\psi (0)\rangle = c_{+}|E_{+}(0)\rangle + c_{-}|E_{-}(0)
\rangle$. Eq.~(\ref{offdiag}) gives the matrix elements of the final density
matrix. The diagonal elements are real-valued, and thus contain no
information about the AGP. The off-diagonal elements, however, do depend on 
the AGP through $\Gamma_{a}(+,-) = \Gamma_{a}(+) - \Gamma_{a}(-)$, where
$\Gamma_{a}(\pm )$ are defined below eq.~(\ref{fstate}). In an NMR experiment,
observation of the AGP is carried out by measuring the transverse magnetization
whose expectation value depends on $\overline{\rho_{+-}(t_{f})}$ 
\cite{sut,fg4}. As the analysis of Section~1 showed,
sufficiently noisy control causes $\overline{\rho_{+-}(t_{f})}$ to
effectively vanish so that all physical consequences of the AGP thus become
unobservable. The stochastic phase shifts $\Gamma_{s}(\pm )$ generated by
noisy control dephase the final superposition state, reducing it to a mixture 
of the states $|E_{\pm}(0)\rangle$
in which all AGP effects are absent. In the NMR setting, the precession of
$\bfB_{a}(t)$ is produced by varying the phase of the rf magnetic field
$\bfB_{rf}(t)=B_{rf}(t)\hat{\mathbf{x}}$. If we imagine the noise is due to rf 
power fluctuations, $H_{s}(t) = -(\gamma\hbar /2)B_{n}(t)\sigma_{x}$ which
identifies $\hat{\mathbf{O}}=\hat{\mathbf{x}}\sigma_{x}$. One can show 
\cite{fg3} that for this type of noise $I_{+-} = 4\tau_{c}T\sin^{2}\theta_{0}$.
If we denote the static magnetic field that splits the nuclear energy levels
by $\bfB_{0} = B_{0}\hat{\mathbf{z}}$, it is well-known \cite{sut} that 
$\sin^{2}\theta_{0} = B_{rf}^{2}/(B_{rf}^{2} + (B_{0}-B_{rf})^{2})$. Using
this result for $I_{+-}$ in eq.~(\ref{deco2}) gives:
\begin{eqnarray}
  \frac{\eta\gamma^{2}T}{4\pi^{2}}
    \left[\frac{B_{rf}^{2}}{B_{rf}^{2}+(B_{0}-B_{rf})^{2}}\right]
     \left(\frac{\overline{P}}{V}\right)
      \left(\frac{\tau_{c}}{\Delta\omega}\right) & \sim & 1 . 
   \label{deco3}
\end{eqnarray} 
Eq.~(\ref{deco3}) identifies the noise properties that impact decoherence, 
and allows a quantitative assessment of how much noise can be tolerated
before all AGP effects are dephased by noise. We believe that this is the 
\textit{first
demonstration of how noisy control causes a dephasing of the AGP}, and the
\textit{first quantitative analysis of the onset condition for this dephasing 
mechanism}. 

\textit{3. Shor Algorithm on a Noisy GQC:} We now consider how noisy control
impacts the performance of a GQC. The universal set $U$ of quantum 
gates used to construct a GQC contains the 1-qubit Hadamard gate $H$ and the
set of all possible 2-qubit controlled-phase gates $B(\phi )$ ($\phi\in [0,
2\pi )$). The action of $B(\phi )$ on the 2-qubit computational basis
states (CBS) $|xy\rangle$ is: $B(\phi ) |xy\rangle = \exp [ixy
\phi ]|xy\rangle$, where x, y = 0, 1. The AGP is used to encode the 
conditional phase shift
$\phi$. We focus on the operation of $B(\phi )$ in the presence of noise as it
is the only gate in the set $U$ that can introduce entanglement into the
dynamics of a GQC. The conditional phase shift $\phi$ is implemented using a
four part pulse sequence $P=P_{0}P_{1}P_{2}P_{3}$, with $P_{0}$ ($P_{3}$) 
applied first (last). Here $P_{0} = C\pi_{1}$; $P_{1}=\overline{C}\pi_{2}$;
$P_{2}=P_{0}$; and $P_{3}=P_{1}$, where $C$ is the cyclic evolution of 
$\bfB_{a}(t)$ introduced in Section~2; $\overline{C}$ is the time-reverse of
$C$; and $\pi_{1}$ ($\pi_{2}$) is a $\pi$-pulse applied to the first
(second) qubit. By appropriate choice of $C$, any phase shift $\phi$ can be
produced. The 1-qubit CBS are: $|0\rangle = |E_{-}(t)
\rangle$ and $|1\rangle = |E_{+}(t)\rangle$. Let $k= (i_{1}, \: i_{2})$ 
with $i_{1}$, $i_{2}$ = 0, 1; then $|E_{k}(t_{0})\rangle
\equiv|E_{i_{1}}(t_{0})\rangle\otimes |E_{i_{2}}(t_{0})\rangle$. The pulse
sequence $P$ maps $k\stackrel{P_{0}}{\rightarrow} k_{1}
\stackrel{P_{1}}{\rightarrow} k_{2}\stackrel{P_{2}}{\rightarrow} k_{3}
\stackrel{P_{3}}{\rightarrow} k_{4}$, with $k_{j} = (i_{1}\oplus
\sum_{l=1}^{j}l \bmod 2,\;
i_{2}\oplus\sum_{l=0}^{j-1}l\bmod 2)$, and $\oplus$ is addition modulo 2.
For $|\psi (t_{0})\rangle =
|E_{k}(t_{0})\rangle$, eq.~(\ref{fstate}) gives the final state $|\psi (t_{f})
\rangle$ with $\Gamma_{a}(k) = (1/\hbar )\sum_{l=0}^{3}\Gamma_{a}(k_{l})$;
$\Gamma_{s}(k) = (1/\hbar )\sum_{l=0}^{3}\Gamma_{s}(k_{l})$; and
$\Gamma_{a}(k_{l}) = \int_{t_{0}(l)}^{t_{f}(l)} dt [E_{k_{l}}(t) - 
\hbar\dot{\gamma}_{k_{l}}(t)]$; $\Gamma_{s}(k_{l}) = \int_{t_{0}(l)}^{t_{f}(l)}
dt H_{s}^{k_{l}k_{l}}(t)$, with $t_{0}(l) = t_{0}+lT$ and $t_{f}(l) = t_{0}(l)
+ T$. For the initial superposition state $|\psi (t_{0})\rangle = 
c_{k}|E_{k}(t_{0})\rangle + c_{j}|E_{j}(t_{0})\rangle$, the final density
matrix 
$\overline{\rho_{kj}(t_{f})}$ is given by eq.~(\ref{offdiag}). The decoherence
factor $\overline{D({k,j})}$ is given by eq.~(\ref{Djk}) with $\Gamma_{s}(k,j)
= \sum_{l=0}^{3}\Gamma_{s}^{l}(k,j)$, and $\Gamma_{s}^{l}(k,j) = 
\int_{t_{0}(l)}^{t_{f}(l)} dt [ H_{s}^{k_{l}k_{l}}(t) - 
H_{s}^{j_{l}j_{l}}(t)]$. Because $\tau_{c}\ll T$ in the adiabatic limit, the
$\Gamma_{s}^{l}(k,j)$ are uncorrelated so that $\overline{\Gamma^{2}_{s}(k,j)}
= \sum_{l=0}^{3}\overline{(\Gamma_{s}^{l}(k,j))^{2}}$. If the 
$\Gamma_{s}^{l}(k,j)$ are also statistically independent, one can show 
\cite{fg3} that $\overline{D(k,j)} = \exp [-\overline{\Gamma_{s}^{2}(k,j)}/2]$
and $\overline{\Gamma_{s}^{2}(k,j)}$ is given by eq.~(\ref{variance}) with 
$I_{kj}\rightarrow \sum_{l=0}^{3}I_{kj}^{l}$. Here $I_{kj}^{l}$ has the same 
form as $I_{kj}$, except that $\hat{\mathbf{O}}_{kj}(t)\rightarrow
\hat{\mathbf{O}}_{kj}^{l}(t) = \hat{\mathbf{O}}_{k_{l}k_{l}}(t) -
\hat{\mathbf{O}}_{j_{l}j_{l}}(t)$. The condition for onset of decoherence
is again 
$\overline{\Gamma_{s}^{2}(k,j)}\sim 4\pi^{2}$. As an example, imagine we input 
the Bell state $(1/\sqrt{2}) \left[ |00\rangle + |11\rangle\right]$ into 
$B(\phi )$. Then $c_{k}=c_{j}= 1/\sqrt{2}$; 
$k = (00)$; and $j = (11)$. If we assume again that the noise is due to
rf power fluctuations, one can show \cite{fg3} that the noise-averaged
entanglement fidelity $\overline{F} = \overline{\langle\psi (t_{0})|\rho
(t_{f})|\psi (t_{0})\rangle} = \frac{1}{2} + \frac{1}{2}\cos\Gamma_{a}(k,j) 
\overline{D(k,j)}$, with $\Gamma_{a}(k,j) = \Gamma_{a}(k)-\Gamma_{a}(j)$. 
Since $\overline{D(k,j)}=\exp [- \overline{\Gamma_{s}^{2}(k,j)}/2]$, 
eq.~(\ref{variance}) shows that it goes smoothly to zero with increasing 
noise variance $\sigma^{2}$. Thus $\overline{F} \rightarrow 1/2$ which is the 
entanglement fidelity for an initial state which is a uniform mixture of the 
states $|00\rangle$ and $|11\rangle$, indicating that entanglement is 
destroyed once the control field driving $B(\phi )$ becomes 
sufficiently noisy. One can show \cite{fg3} that for this Bell state 
$\sum_{l=0}^{3}I^{l}_{kj}= 32\tau_{c}T\sin^{2}\theta_{0}$ so that the 
condition for onset of decoherence (and loss of entanglement) is
$(2\eta /\pi^{2})\, (\gamma^{2}/\Delta\omega)\,
 (\overline{P}/V)\, (\tau_{c}T\sin^{2}\theta_{0})\,\sim 1$.
Recent work \cite{t+n} also found that noise will severely impact the
performance of $B(\phi )$. 

We now show how a GQC containing noisy controlled-phase gates impacts the 
computational efficiency
of the Shor algorithm for factoring an integer $N$ \cite{shr}. Number 
theoretic arguments reduce factoring to finding the period $r$ of the function
$F(a) = y^{a}\bmod N$, where $y$ is co-prime with $N$. The algorithm begins
by preparing the GQC in the state
$|\psi_{0}\rangle = (1/\sqrt{A+1})\sum_{j=0}^{A}|jr + l \rangle$. Here $r$ is
the period of $F(a)$; $l$ is the result of a measurement carried out during
preparation of $|\psi_{0}\rangle$; and $A$ is the largest integer such that
$Ar+l<q$, where $q$ is chosen such that $N^{2}\leq q\leq 2N^{2}$. The 
algorithm implements a discrete Fourier transform modulo $q$ 
($\mathnormal{DFT}_{q}$) on $|\psi_{0}\rangle$: $|\psi_{1}\rangle =
\mathnormal{DFT}_{q}|\psi_{0}\rangle = \sum_{c=0}^{q-1}\tilde{f}(c)|c\rangle$,
where
\begin{equation}
\tilde{f}(c) = \frac{\sqrt{r}}{q}\,\sum_{j=0}^{q/r-1}\exp\left[
            \frac{2\pi i}{q}(jr+l)c + i\Gamma_{s}(c)\right] .
\label{famp}
\end{equation}
The $\mathnormal{DFT}_{q}$ requires the application of $L(L-1)/2$ 
controlled-phase gates, where $L\equiv\log_{2}q$. Noisy control of this
gate causes the
stochastic phase shift $\Gamma_{s}(c)$ to appear in eq.~(\ref{famp}). 
One can show \cite{fg3} that $\overline{\Gamma_{s}(c)}=0$, and the variance 
$\overline{\Gamma_{s}^{2}(c)}$ is given by eq.~(\ref{variance}) with 
$\eta = L(L-1)/2$; $\sigma^{2} = \overline{P}/V\Delta\omega$; and
$I_{kj}\rightarrow \sum_{l=0}^{3}I_{kj}^{l}$. The final step in the algorithm
is a measurement which produces the result $c$ with noise-averaged 
probability $P(c)=\overline{|\tilde{f}(c)|^{2}}$. 
In the absence of noise, constructive interference in 
eq.~(\ref{famp}) occurs when $|rc-c^{\prime}q|\leq r/2$, which determines
a unique $c^{\prime}$, and consequently, $P(c^{\prime})=P(c)$. The measurement 
result
$c$ yields the period $r$ only if $c^{\prime}$ and $r$ are co-prime. The
success probability for the algorithm is then 
$P_{suc}=\sum_{c^{\prime}}^{\prime}
P(c^{\prime})$, where in the primed sum only those $c^{\prime}$ appear
that are both less than, and co-prime with, $r$. One can show \cite{fg3} that:
\begin{eqnarray}
\lefteqn{\hspace{-0.25in}\overline{|\tilde{f}(c)|^{2}} \hspace{0.05in} = 
  \hspace{0.1in} \frac{1}{q}} \nonumber \\
  &  &  \hspace{-0.25in}+ \frac{2r}{q^{2}} \sum_{k=0}^{q/r-1}\sum_{j>k}
                            \cos\left[
                              \frac{2\pi (j-k)}{q}(rc \bmod q)\right]
                               \overline{D(k,j)} ,
\label{fnoyz}
\end{eqnarray}
where $\overline{D(k,j)}$ is the decoherence factor due to noise. As above,
$\overline{D(k,j)} = \exp\left[ 
-\overline{\Gamma_{s}^{2}(k,j)}/2\right]$ which, by eq.~(\ref{variance}),
goes smoothly to zero as the noise variance $\sigma^{2}$ increases. In the
decoherence limit only the first term on the RHS of eq.~(\ref{fnoyz}) 
survives so that $\overline{|\tilde{f}(c)|^{2}}\sim 1/q\sim 1/N^{2}$.
This result is independent of
$c$ so that $P_{suc}\sim (1/N^{2})\phi (r)$, where $\phi (r)$ is Euler's Phi
function which gives the number of integers that are less than and co-prime 
with $r$. For large $N$, $r\alt N$ and $\phi (r) \sim r/\log r\sim N/\log N$.
Thus, in the decoherence limit, $P_{suc}\sim 1/(N\log N) = 2^{-\log N}/\log N$.
We see that the algorithm must be run on average $2^{\log N}$ times to
obtain $P_{suc}\sim 1$. This is exponential in the problem size $\log N$, 
indicating that sufficiently strong noise destroys the computational 
efficiency of the Shor algorithm. This contrasts with the (computationally
efficient) noiseless Shor algorithm whose runtime scales linearly with 
$\log N$. Onset of decoherence occurs when
$\overline{\Gamma_{s}^{2}(c)}\sim 4\pi^{2}$, and for a GQC, $\sum_{l=0}^{3}
I_{kj}^{l}\sim \tau_{c}T\sin^{2}\theta_{0}$ so that decoherence occurs when:
\begin{equation}
\left(\frac{\overline{P}}{V}\right)\,\left(\frac{\tau_{c}}{\Delta\omega}
 \right) \agt \frac{\pi^{2}}{TL(L-1)\gamma^{2}\sin^{2}\theta_{0}} .
\label{gqcdeco}
\end{equation}
Adiabatic operation of a GQC requires $T$ to be large; and the desire to 
factor large $N$ means $L = \log_{2} q \sim\log_{2}N$ will also be large. 
Thus the RHS of eq.~(\ref{gqcdeco}) is expected to be small, so that managing 
noise will in fact be a significant 
issue for a GQC after all. \textit{Eq.~(\ref{gqcdeco}) allows a quantitative
estimate of how much noise can be tolerated by a GQC before decoherence
due to noisy control undermines its operation. We believe that the above
analysis is the first to quantitatively demonstrate how noisy control 
destroys the efficiency of Shor algorithm when run on a sufficiently
noisy GQC}.

\begin{acknowledgments}
I would like to thank T. Howell III for continued support; the National
Science Foundation for support through grant NSF-PHY-0112335;
and the Army Research Office for support through grant
DAAD19-02-1-0051.
\end{acknowledgments}

\end{document}